\documentstyle[12pt]{article}
\title{{\normalsize
\begin{flushright}SUSX-TH-95/71\\
{\tt hep-ph/9503457}\\
March 1995\\
\end{flushright}}
\vspace{1 cm}
Bound States and Instabilities of Vortices}
\author{Michael Goodband\thanks{e-mail:
{\tt m.goodband@sussex.ac.uk }}
and Mark Hindmarsh\thanks{e-mail: {\tt m.b.hindmarsh@sussex.ac.uk }}
}
\date{}

\begin{document}
\maketitle
\vspace{-24pt}
\begin{center}
{\it
 School of Mathematical and Physical Sciences\\
University of Sussex\\
Brighton BN1 9QH\\
UK
}
\end{center}

\vfill

\begin{abstract}

We examine the spectrum of small perturbations around global
and local (gauge) abelian vortices, using simple numerical matrix
techniques.  The results are of interest both for cosmic strings
and for their condensed matter analogues, superfluid and
superconductor vortices.  We tabulate the instabilities of
higher winding number vortices, and find several bound states.
These localised coherent oscillations of the order parameter
can be thought of as particle states trapped in the core of the
string.

\end{abstract}
\newpage

\section*{Introduction}

The ideas of gauge unification and spontaneous symmetry breaking lie
at the heart of modern particle physics.
The well established view is of some Grand Unified Theory with
gauge symmetry group $G$, suffering a series of
spontaneous symmetry breakings which reduce the group to that of
the Standard Model. In a cosmological
context this series of spontaneous symmetry breakings is seen
to occur as the universe cools. The first
symmetry breaking occurs at an energy of the order $ 10^{16}$ Gev,
and it is possible that topological defects
may form, via the Kibble mechanism, at this transition
\cite{Kib76,VilShe94}.
Such objects
have many important implications: for example, defects of
dimension 2 (domain walls) and 0 (monopoles) would
come to dominate the energy density of the Universe soon
after a GUT phase transition, which would necessitate a
period of inflation if the theory admitted such defects.

When a U(1) symmetry is broken at a phase transition, there
arises a network of cosmic strings \cite{HinKib94},
defects of dimension 1, which may
provide seeds for large scale structure formation in the universe.
Because strings are very thin and the
cosmological scales of interest are many orders of magnitude bigger,
they have been approximated by line sources
with energy-momentum density $ diag(\mu ,0,0,-\mu ) $ where $ \mu $
is the energy density of the classical
solution. There are, however, a few instances when one is interested in
the
string structure. They include:  calculating scattering cross-sections
(which are important for baryon number violation by strings
\cite{Per+91}); finding bound states;
quantum corrections to the energy; and curvature corrections for
highly curved string \cite{Curve}. All of these
require a knowledge of the perturbation spectrum about the string.

In this paper we find the perturbation spectra in the backgrounds
of global and local U(1) vortices.  Simple U(1) strings provide
an ideal test-bed for developing numerical techniques for eigenvalues.
They are not trivial, like the kink \cite{Raj82}, nor are they
as geometrically involved as 3D objects such as the sphaleron
\cite{Sph}.
However, one has to deal with all the same problems: gauge choice
and numerical stability being the principal ones.  Furthermore,
they are a stepping stone on the way to an efficient technique
for mapping the stability region of the electroweak string,
which we shall describe in a separate publication \cite{GooHin95b}.
Our methods are quite straightforward:  we find the vortex solution
using shooting or relaxation algorithms.  The numerical profile
functions are then used in constructing a discretised matrix of
second derivatives of the fields.  This matrix is then diagonalised
to find the eigenvalues and eigenfunctions, using the standard
linear algebra packages incorporated into {\sc matlab}.

We find a
continuum of scattering states, oscillatory at infinity, for
each particle species, obeying the expected dispersion relation
$\omega^2 = {\bf k}^2 + m^2$.  There are unstable modes for
higher winding number strings, corresponding to strings splitting
in various ways. We tabulate these, and show how different angular
momenta in the modes lead to different splitting configurations.
We also find bound states: $N$ on
global strings of winding number $N$ (we searched up to $N=5$);
but on the gauge strings we found only one, which crossed
into the continuum for a sufficiently large ratio of the
scalar to gauge coupling.  The bound states are excitations of
the fields at the core of the string, decaying exponentially at
infinity: one can think of them as particles trapped on the string.

As is well known, global strings are closely related to superfluid
vortices, and local ones to flux tubes in superconductors.  We conclude
by discussing the connection and outlining the possible significance
of our results for condensed matter physics.

\section{Perturbations about global strings}

The model considered in this section has a complex scalar field with the
lagrangian density
\begin{displaymath}
{\cal L} = ( \partial^{\mu } \phi )^{\ast } ( \partial_{\mu } \phi ) -
\lambda
(\phi^{\ast} \phi -  \eta^{2} /2 )^{2},
\end{displaymath}
invariant under global U(1) transformations. The field acquires a vacuum
expectation value,
breaking the symmetry and giving rise to a particle spectrum of one
Goldstone
boson and a
Higgs boson of mass $ \sqrt{2 \lambda \eta^{2} } $.

As is well known, there exist time independent classical vortex
solutions to
the equation
of motion
\begin{displaymath}
- \partial^{2} \phi + 2\lambda (| \phi |^{2} - \eta^{2} /2)\phi = 0.
\end{displaymath}
The solutions are {\it z}-independent and are of the form $ \phi = (\eta /\surd
2)
f(\rho )
e^{i N \theta } $ where $ N $
is the winding number. The energy per unit length of such configurations
is
given by
\begin{displaymath}
E = \pi \eta^{2} \int \rho d\rho \left( \left( \frac{df}{d\rho }
\right)^{2}
+ \frac{N^{2} }{\rho^{2} } f^{2}
+ \frac{\lambda \eta^{2} }{2} (f^{2} -1 )^{2} \right).
\end{displaymath}
The energy is logarithmically divergent for large $ \rho $
\begin{displaymath}
E = E_{core} + \pi \eta^{2} N^{2} \ln \left( \frac{R }{\rho_{c} }
\right),
\end{displaymath}
where $ \rho_{c} $ is the core radius.  In physical situations there
exists a physical cut off, such as a container or a nearby anti-vortex,
making the energy finite.  It can be seen from the $ N^{2} \ln (R ) $
behaviour that two $ N=1 $ global strings have significantly lower
energy than one $ N=2 $ global string.  The decay of an $ N=2 $ string
to two $ N=1 $ strings conserves the topological charge and so it would
be expected that the decay would occur.  Similarly, one expects the decay
of higher $ N $ strings into strings of lower winding number, provided the
decay conserves the topological charge.  Decays which would appear to be
possible on energetic grounds (eg.  one $ N=2 $ string to one $ N=1 $
string) but do not conserve the topological charge, do not occur because
there is an infinite energy barrier seperating the two configurations.
So only the $ N=1 $ string is expected to be stable.

Now consider {\it z}-independent perturbations to the classical string
background $ \phi_{c} (x)  $ of the form $ \phi_{c}
({x^i} ) + \epsilon \delta \phi ({x^i} ) e^{-i\omega t} $ and $
\phi_{c}^{\ast } ({x^i} ) + \epsilon \delta \phi^{\ast } ({x^i} )
e^{i\omega t}$ where $i=1,2$.  This gives corrections to the energy
\begin{displaymath}
E = E(\phi_{c} ) + \epsilon^{2} E_{2} + O(\epsilon^{3} ),
\end{displaymath}
where
\begin{displaymath}
E_{2} =  \int d^{2} x (\delta \phi^{\ast } ,\delta \phi ) M \left(
\begin{array}{c}
  \delta \phi \\ \delta \phi^{\ast }
  \end{array} \right).
\end{displaymath}
The matrix $ M $ is the perturbation operator, obtained by taking second
functional derivatives of the energy.  To find the perturbative modes
about the string, which can be interpreted as particle states, we solve
the coupled eigenvalue problem
\begin{displaymath}
M \left( \begin{array}{c}
           \delta \phi \\ \delta \phi^{\ast }
           \end{array}
   \right) = \omega^{2} \left(
   \begin{array}{c}
       \delta \phi \\ \delta \phi^{\ast }
    \end{array}
\right),
\end{displaymath}
where
\begin{displaymath}
M = \left(
\begin{array}{cc}
 - \nabla^{2} + 2\lambda (2 | \phi_{c} |^{2} - \eta^{2} /2 )  & 2\lambda
 {\phi_{c}^{\ast }}^2  \\
 2\lambda \phi_{c}^{2} & - \nabla^{2} + 2\lambda (2 | \phi_{c} |^{2} -
\eta^{2} /2 )
\end{array}
\right).
\end{displaymath}
The string is a cylindrically symmetric solution, so it is convenient to
expand
the perturbations in angular
momentum states
\begin{displaymath}
\delta \phi = \sum_{m} s_{m} e^{i(N+m) \theta } \hspace{8mm} \mbox{and}
\hspace{8mm}
\delta \phi^{\ast } = \sum_{-m} s_{-m}^{\ast } e^{-i(N-m) \theta }.
\end{displaymath}
It is also useful to rescale the coordinates by defining $ r = \sqrt{2
\lambda
\eta^{2} } \rho $ to remove the
parameters $ \lambda  $ and $ \eta  $ from the equations. So rescaling
and
substituting in the string solution
and perturbation expansion gives the coupled eigenvalue equations for
the
functions $ s_{m} $ and $ s_{-m}^{\ast } $
\begin{eqnarray}
  \left( - \nabla_{r}^{2} + \frac{(m-N)^{2} }{r^{2} } +
  \frac{1}{2} (2 f^{2} - 1 ) \right) s_{-m}^{\ast }
   + \frac{1}{2} f^{2} s_{m} & = & \omega^{2} s_{-m}^{\ast } \nonumber
\\
  \left( - \nabla_{r}^{2} + \frac{(m+N)^{2} }{r^{2} } +
  \frac{1}{2} (2 f^{2} - 1 ) \right) s_{m}
   + \frac{1}{2} f^{2} s_{-m}^{\ast } & = & \omega^{2} s_{m},  \nonumber
\end{eqnarray}
where $\omega^2$ is now dimensionless.  To recover the dimensionful
values, one should multiply by $m_H^2= 2\lambda\eta^2$.
Note that the eigenvalue equations are the same after complex
conjugation
and changing $ m \rightarrow -m $,
so we need consider only $ m \geq 0 $.

If we set $( s_{m} $,$ s_{-m}^{\ast } )$ to real and imaginary parts,
we can seperate the above eigenproblem into
two separate eigenproblems with explicitly real fields
\begin{eqnarray}
\left( \begin{array}{cc}
        D_{1} + \frac{1}{2} (2f^{2} - 1) & \frac{1}{2} f^{2} \\
        \frac{1}{2} f^{2} & D_{2} + \frac{1}{2} (2f^{2} - 1)
        \end{array}
        \right)
        \left(
        \begin{array}{c}
          s_{m}^{r} \\ s_{-m}^{r}
        \end{array}
        \right)
        & = & \omega^{2} \left(
        \begin{array}{c}
          s_{m}^{r} \\ s_{-m}^{r}
        \end{array}
        \right) \\
\left( \begin{array}{cc}
        D_{1} + \frac{1}{2} (2f^{2} - 1) & \frac{1}{2} f^{2} \\
        \frac{1}{2} f^{2} & D_{2} + \frac{1}{2} (2f^{2} - 1)
        \end{array}
        \right)
        \left(
        \begin{array}{c}
         s_{m}^{i} \\ -s_{-m}^{i}
        \end{array}
        \right)
        & = & \omega^{2} \left(
        \begin{array}{c}
         s_{m}^{i} \\ -s_{-m}^{i}
        \end{array}
        \right)
\end{eqnarray}
where $ D_{1} = -\partial_{r}^{2} + (m-N)^{2} /r^{2}  $ and
$ D_{2} = -\partial_{r}^{2} + (m+N)^{2} /r^{2} $.

First consider the vacuum where $ f=1 $ and $ N=0 $.
Then we have $ D_{1} = D_{2} $ and we can diagonalise
$ M $ with the with the eigenvectors $ ( 1 ,1 )^{T} s_{m} $
and $ (1 ,-1)^{T} s_{m} $ to give
\begin{eqnarray}
  \left( -\nabla_{r}^{2} + \frac{m^{2} }{r^{2} } + 1 \right) s_{0}
  & = & \omega^{2} s_{0} \\
  \left( -\nabla_{r}^{2} + \frac{m^{2} }{r^{2} } \right) s_{0}
  & = & \omega^{2} s_{0}
\end{eqnarray}
We can see that for the vacuum the solutions are Bessel's functions $
J_{m} (kr) $ where for (3) $ k $ is given by the continuum $ \omega^{2}
= k^{2} + 1 $ and for (4) by $ \omega^{2} = k^{2} $.  The eigenvalue
spectrum of (3) corresponds to the spectrum a Higgs particle with rest
mass 1 in units of $ 2 \lambda \eta^{2} $, whereas the eigenvalue
spectrum from (4) has a continuum down to $ \omega^{2} = 0 $
corresponding to Goldstone modes.

\subsection*{Numerical method}

The eigenproblem above was solved with the boundary conditions $ s_{m}
$,$ s_{-m} \rightarrow 0 $ as $ r \rightarrow \infty $ for various
values of $ m $.  The profile for a string of winding $ N $ was found by
solving the equation for $ f $ using a shooting method.  This profile
was then substituted into the discretized perturbation operator $ M $
and the eigenvalue problem was then solved by the standard matrix
methods used by {\sc matlab}.  This proceedure was performed for
linearly discretized lattices with 64, 128 and 256 points.  The results
for the 256 point lattice are shown in Tables 1 and 2.  The numerical values
for the zero modes give an estimate of the accuracy of the method.
The eigenvalue problem was also solved for a linear discretization in the
variable $ \rho $ where $ \rho = \tanh (r) $.  This gives a non-linear
discretization in $ r $ with more points in the core of the string.  The
results from this are in agreement with those quoted in Table 1 to
the error given by the values for the zero modes.

\subsection*{Zero modes}

The string is a solution centered at a point in a two dimensional plane
and so there must be a perturbed field
configuration corresponding to the infinitesimal translation
\begin{eqnarray}
\phi (x) \rightarrow \phi (x + \delta x) & = & \phi (x) +
\frac{\partial \phi }{\partial x^{i}} \delta x^{i}  \nonumber \\
      & = & \phi + \epsilon \delta \phi \nonumber
\end{eqnarray}
For example, for the translation mode in the {\it x}-direction we take
the
{\it x}-derivative of the string solution
\begin{eqnarray}
\frac{\partial \phi_{c} }{\partial x} & = & \cos \theta \frac{\partial
\phi_{c} }{\partial r}
- \frac{\sin \theta }{r} \frac{\partial \phi_{c} }{\partial \theta}
\nonumber \\
  & = & \cos \theta \frac{df}{dr} e^{iN \theta } -
  \frac{iN \sin \theta }{r} f e^{iN \theta }  \nonumber \\
  & = & \frac{e^{iN\theta } }{2} \left( \left( \frac{df}{dr} -
  \frac{Nf}{r} \right) e^{i\theta }
  + \left( \frac{df}{dr} + \frac{Nf}{r} \right) e^{-i\theta } \right).
  \nonumber
\end{eqnarray}
As can be seen, the translation mode has $ m=1 $, and the forms for
$ s_{1} , s_{-1} $ can be read off from
above and shown to satisfy
\begin{eqnarray}
M \left( \begin{array}{c}
      \delta \phi \nonumber \\
      \delta \phi^{\ast} \nonumber
      \end{array} \right) = 0.
\end{eqnarray}
The numerical results for this zero mode are shown in the table and, as
mentioned above, they give an estimate of the error in the eigenvalues.

\subsection*{Bound states}

For the case $ m=0 $, we have $ D_{1} = D_{2} $ and again we can
diagonalise $ M $ with the eigenvectors
$ ( 1 ,1 )^{T} s_{0} $ and $ (1 ,-1)^{T} s_{0} $ to give
\begin{eqnarray}
  \left( -\nabla_{r}^{2} + \frac{N^{2} }{r^{2} } + \frac{1}{2} (3f^{2}
-1)
  \right) s_{0} & = & \omega^{2} s_{0} \\
  \left( -\nabla_{r}^{2} + \frac{N^{2} }{r^{2} } + \frac{1}{2} (f^{2}
-1)
  \right) s_{0} & = & \omega^{2} s_{0}.
\end{eqnarray}

The eigenvalue spectra of the above equations have continua as for the
vacuum, plus the appearance of descrete eigenvalues with $ \omega^{2} <
1 $ for (5) due to the potential confining states on the string.  These
later states can be interpreted as Higgs particles trapped on the string
or as oscillations of the string thickness.  The continuum states are
scattering states with solutions tending to phase shifted Bessel's
functions as $r\to\infty$.  We found that there $ N $ bound states for a
string of winding number $ N $ for $ N=1, \ldots ,5 $.  The solutions to
(6) consist solely of scattering states.

If we consider the perturbations as a function of $ z $ then the
perturbation expansion is
\begin{displaymath}
\delta \phi = \sum_{m} s_{m} e^{i(N+m) \theta } e^{ik_{z} z}
\hspace{8mm} \mbox{and} \hspace{8mm}
\delta \phi^{\ast } = \sum_{-m} s_{-m}^{\ast } e^{-i(N-m) \theta }
e^{-ik_{z} z}.
\end{displaymath}
and the solutions to the corresponding eigenvalue problem are similar to
those given above but with $ \omega^{2} =
k_{r}^{2} + k_{z}^{2} + 1 $ etc. The bound modes given above can then be
interpreted as travelling modes along the string. The profile for
the bound mode of the $N=1$ vortex is shown in Figure 1 together with the
string profile.

\subsection*{Decay modes}

As noted earlier all global strings with $ N \geq  2 $ are expected
to decay to strings
of lower winding number. The modes corresponding to these decays
occur for $ m \geq 2 $. The eigenvalues are listed in Table
1, and we display $\phi + \epsilon\delta\phi$ (with a
large $\epsilon$) in Figures 2--5, to give a graphical
representaion of the decay.

For $ N=2 $ there is an $ m=2 $ mode with negative eigenvalue,
corresponding to the string splitting into two
$ N=1 $ strings (Figure 2).

For $ N=3 $ there is an $ m=2 $ mode with negative eigenvalue
corresponding to the string splitting into three
$ N=1 $ strings in a line (Figure 3). The $ m=3 $ negative mode
corresponds to the string splitting into three $ N=1 $
strings in a triangular configuration (Figure 4). There is also an
$ m=4 $ negative mode which gives rise to five vortex
cores (Figure 5). To see what is happening here consider the small
$ r $ behaviour of the perturbed string solution
\begin{eqnarray}
  \phi & = & \phi_{c} + \delta \phi \nonumber \\
  \phi & = & f e^{i3\theta } + s_{4} e^{i7\theta } + s_{-4} e^{-i\theta
},  \nonumber
\end{eqnarray}
which for small $r$ is given by
\begin{displaymath}
 \phi \sim r^{3} e^{i3\theta } + r^{7} e^{i7\theta } + r e^{-i\theta },
\end{displaymath}
so can see that $ \phi \rightarrow  r e^{-i\theta } $ as $ r \rightarrow
0 $. There is therefore the beginnings of
an anti-vortex at the origin. Thus the net winding number is three,
which is just as well since it is a conserved
topological charge.

This is the first example of a more general feature that strings of
winding number $ N $ possess negative modes
for $ 2 \leq m \leq 2 N - 2 $ where the string splits into strings of
lower winding number, $ m + 1 $ of them for
$ m > N $ and for $ m=N $  the string splits into $ N $ strings of unit
winding. If we consider the energy of the
string and the conservation of topological charge, the general rule for
the allowed decays is that an
$ N $ wound string $ \rightarrow M $ strings of winding $ M_{i} $ , such
that
\begin{eqnarray}
     \sum_{i} M_{i} & = & N \\
     \sum_{i} M_{i}^{2} & < & N^{2}
\end{eqnarray}
The first condition is the conservation of topological charge. The
second condition comes from the $ N^{2} \ln (R)$
dependence of the energy for large $ R $. The $ \sum_{i} M_{i}^{2} =
N^{2} $ possibility is ruled out due to the
dependence of the core energy on $ N $. If the energy of the string is
divided up as
\begin{eqnarray}
E_{Goldstone} & = & \pi \eta^{2} N^{2} \int f^{2} \frac{dr}{r}
\nonumber \\
E_{Higgs} & = & \pi \eta^{2} \int r dr \left( \left( \frac{df}{dr}
\right)^{2} + \frac{1}{4}(f^{2} - 1)^{2} \right) \nonumber
\end{eqnarray}
then $ E_{Higgs} $ is finite whereas $ E_{Goldstone} $ has the $ \ln (R)
$ dependence for large $ R $. As can be
seen from Table 1, the sum of $ E_{Higgs} $ for the $ M $ strings is
greater than that of
for the string of winding $ N $.

Although there are decay modes which begin to generate anti-vortices at
the core for small perturbations, do they remain
as the string decays further? In any physical situation this is
unlikely. This is because the decay modes with anti-vortices
have a number of unit winding number vortices arranged arround them in a
symmetrical arrangement (for instance the $ m=6 $
decay mode of the $ N=5 $ string has a unit winding number anti-vortex
at the center and six unit winding number vortices
arranged arround it in a hexagon), and any slight asymmetry will result
in the anti-vortex being attracted to one of the
vortices and annihilating.

\section{Perturbations about local strings}

In this section we are considering the abelian-Higgs model with the
lagragian density
\begin{displaymath}
{\cal L} = - \frac{1}{4} F_{\mu \nu } F^{\mu \nu } + (D_{\mu
} \phi )^{\ast }
(D^{\mu } \phi ) - \lambda (| \phi |^{2} - \eta^{2} /2 )^{2},
\end{displaymath}
where
\begin{eqnarray}
D_{\mu } & = & ( \partial_{\mu } + ie A_{\mu } ) \phi \nonumber  \\
F_{\mu \nu } & = & \partial_{\mu } A_{\nu } - \partial_{\nu } A_{\mu }.
\nonumber
\end{eqnarray}
The lagrangian is invariant under local U(1) transformations. This
symmetry is broken by the Higgs field
acquiring a vacuum expectation value, and this gives a particle spectrum
of a massive gauge boson of mass
$ e \eta $ and a massive Higgs boson of mass $ \sqrt{2 \lambda \eta^{2}
} $. As shown by Nielsen and Olesen \cite{N.O.}
there exist time independent classical vortex solutions to the time
independent field equations
\begin{eqnarray}
D_{a} D_{a} \phi & = & 2\lambda ( | \phi |^{2} - \eta^{2} /2) \phi
\nonumber \\
\partial_{a} F_{ab} & = & -ie (\phi^{\ast } D_{a} \phi - \phi
(D_{a} \phi )^{\ast } ), \nonumber
\end{eqnarray}
where $a=1,2,3$.
The cylindrically symmetric $ z $-independent vortex solutions can be
written in the form $ \phi = (\eta / \surd 2) f(\rho ) e^{iN \theta } ,
A_{\theta } = -Na(\rho )/e\rho  $ and $ A_{\rho } = 0 $ where $ N $ is
the winding number of the string. The energy per unit
length of the configuration is given by
\begin{displaymath}
E = \pi \eta^{2} \int \rho d\rho \left( \left( \frac{df}{d\rho }
\right)^{2} + \frac{N^{2} }{\rho^{2} } f^{2} (1 - a)^{2}
+ \frac{\beta }{2} ( f^{2} -1)^{2} + \frac{N^{2} }{2\rho^{2} } \left(
\frac{da}{d\rho } \right)^{2} \right),
\end{displaymath}
where the new dimensionless coordinates $r$ is defined by
$$ r = \frac{e \eta }{\sqrt{2} } \rho =  \frac{m_{A}
}{\sqrt{2} } \rho,
$$
and the parameter $ \beta $ is
\begin{displaymath}
\beta = \frac{m_{H}^{2} }{m_{A}^{2} } = \frac{2 \lambda }{e^{2} }.
\end{displaymath}
It will also be convenient to rescale the fields by making the
replacements
$$
\phi \to \frac{\phi\eta }{\sqrt{2}} , \qquad A_\mu \to \frac{\eta}{\sqrt
2}A_\mu.
$$
Dimensionless fields and coordinates will be used throughout
this section.
The  stability of $ N>1 $ Nielsen-Olesen vortices to splitting into
vortices of lower winding has been studied
previously \cite{Bogo}, where it was shown that they are stable for $ \beta <1
$
and unstable for $ \beta > 1 $. This can be seen from Figure 6
showing the energy per vortex for three values of $ \beta $. For $
\beta>1 $ the energy per vortex is lowest for
$ N=1 $, and so a vortex with winding number greater than one will split
into vortices of lower winding number
and so move left on the curve. For $ \beta<1 $ the reverse is true.

\subsection*{Perturbations and gauge fixing}

Now, consider perturbations to the classical string solution of the form
$ \phi_{c} (x ) + \epsilon \delta \phi (x ) e^{-i\omega t} $,
$ \phi_{c}^{\ast } (x ) + \epsilon \delta \phi^{\ast } (x )e^{i\omega t}
$
and $ A_{\mu } (x )+ \epsilon \delta A_{\mu } (x )e^{-i\omega t} $.
This gives corrections to the action
\begin{displaymath}
S = S(\phi_{c} , A_{\theta}^{c} ) + \epsilon^{2} S_{2} + O(\epsilon^{3}
),
\end{displaymath}
where
\begin{displaymath}
S_{2} = \frac{1}{2} \int d^{4} x \delta \Phi^\dagger   {\cal D}
\delta \Phi ,
\end{displaymath}
and where $ {\cal D} $ is the perturbation operator and
$ \delta \Phi^{\dagger } = (\delta \phi^{\ast }e^{i\omega t} ,
\delta \phi e^{-i\omega t},\delta A_{\mu } e^{-i\omega t} ) $.
The perturbation operator $ {\cal D} $ is given by
\begin{eqnarray*}
\lefteqn{{\cal D} = } \\
 & & \left(  \begin{array}{c}
    - \partial^{2} - 2iA_{\mu  } \partial^{\mu } + A_{\mu }^2 - \beta
(2| \phi |^{2} -1 ) \hspace{10mm}
    -\beta \phi^{2} \hspace{10mm}
    -2i\partial_{\mu } \phi - i\phi \partial_{\mu } + 2 A_{\mu} \phi  \\
    -\beta \phi^{\ast^{2} } \hspace{10mm}
    - \partial^{2} + 2iA_{\mu } \partial^{\mu } + A_{\mu}^{2} - \beta
(2| \phi |^{2} -1 ) \hspace{10mm}
    2i\partial_{\mu } \phi^{\ast } + i\phi^{\ast } \partial_{\mu } + 2
A_{\mu } \phi^{\ast }  \\
    i\partial_{\mu } \phi^{\ast } - i \phi^{\ast } \partial_{\mu } + 2
A_{\mu } \phi^{\ast } \hspace{4mm}
   -i\partial_{\mu } \phi + i \phi \partial_{\mu } + 2 A_{\mu } \phi
\hspace{4mm}
   g^{\mu \nu} \partial^{2} - \partial_{\mu } \partial_{\nu} + 2 g^{\mu
\nu} | \phi |^{2}
   \end{array}
   \right). \nonumber
\end{eqnarray*}
and the equations of motion for the perturbations are $ {\cal D} \delta
\Phi =0 $.

In order to set up an eigenvalue problem for the perturbations we need
to remove the linear derivative terms,
as well as removing the gauge degrees of freedom. To do this we choose
the background gauge condition
\cite{C-L}
\begin{eqnarray}
F(A) = \partial_{\mu} \delta A^{\mu} - (\delta \phi^{\ast } \phi_{c} -
\phi_{c}^{\ast}
\delta \phi ) = 0.
\end{eqnarray}
which is imposed by adding the gauge fixing term
\begin{displaymath}
{\cal L}_{GF} = \frac{1}{2} | F(A) |^{2}
\end{displaymath}
to the lagrangian. This enables us to seperate out the time derivatives
to give the eigenvalue equations
\begin{displaymath}
M^{GF} \left( \begin{array}{c}
               \delta \phi \\
               \delta \phi^{\ast } \\
               \delta A_{i}
               \end{array}
               \right) = \omega^{2} \left(
               \begin{array}{c}
               \delta \phi \\
               \delta \phi^{\ast } \\
               \delta A_{i}
               \end{array}
               \right)
\end{displaymath}
and
\begin{displaymath}
\left( -\nabla^{2} + 2 |\phi |^2 \right) \left(
           \begin{array}{c}
            \delta A_{0} \\
            \delta A_{z}
            \end{array}
            \right) = \omega^{2}
            \left(
             \begin{array}{c}
             \delta A_{0} \\
             \delta A_{z}
             \end{array}
             \right)
\end{displaymath}
where $ i=1,2 $ and the gauge fixed perturbation operator is
\begin{eqnarray*}
\lefteqn{M^{GF}= } \\
 & & \left(  \begin{array}{c}
   -\nabla^{2} -2iA_{i} \nabla_{i} +A_{i}^{2} +\beta (2| \phi |^{2} -1 )
+ | \phi |^{2} \hspace{6mm}
   ( \beta -1 ) \phi^{2}  \hspace{6mm}
   -2i\nabla_{i} \phi + 2 A_{i} \phi  \\
   (\beta -1) \phi^{\ast^{2} } \hspace{6mm}
   -\nabla^{2} +2iA_{i} \nabla_{i} +A_{i}^{2} +\beta (2| \phi |^{2} -1 )
+ | \phi |^{2} \hspace{6mm}
   2i\nabla_{i} \phi^{\ast } + 2 A_{i} \phi^{\ast }  \\
   2i\nabla_{i} \phi^{\ast } + 2 A_{i} \phi^{\ast } \hspace{17mm}
   -2i\nabla_{i} \phi + 2 A_{i} \phi \hspace{17mm}
   -\delta^{ij} \nabla^{2} + 2 \delta^{ij} | \phi |^{2}
   \end{array}
   \right). \nonumber
\end{eqnarray*}

The $ \delta A_{0} $ and $ \delta A_{z} $ perturbations decouple because
the background string solution is
independent of $ t $ and $ z $.

The gauge condition (9) above does not fix the gauge completely. For
comparison, consider the more familiar
coulomb gauge $ \nabla \cdot A =0 $ ,where it is still possible to make
gauge transformations
\begin{displaymath}
A_{i} \rightarrow A_{i} - \nabla_{i} \Lambda
\end{displaymath}
and the gauge transformed field still satisfy the gauge, as long as $
\Lambda $ is harmonic. Similarly for
the background gauge condition it is still possible to make gauge
transformations for $ \Lambda $ satisfying
\begin{displaymath}
(-\partial^{2} + 2 | \phi |^{2} ) \Lambda = 0 .
\end{displaymath}
This is why in quantum field theory the above gauge fixing term would be
accompanied by a Fadeev-Popov
ghost term of the form
\begin{displaymath}
{\cal L}_{FPG} = \frac{1}{2} \Lambda^{\ast } ( - \partial^{2} + 2 |
\phi_{c} |^{2} ) \Lambda .
\end{displaymath}

In 2+1 dimensions the perturbation operator for $ \delta A_{0} $ and the
ghosts is the same, so $ \delta A_{0} $ and
$ \Lambda  $ will have the same eigenvalue spectrum. In any calculation
of quantum corrections, the perturbation operator for the ghosts
enters with the opposite sign and a factor of $ 2 $ compared to that for
$ \delta A_{0} $. Thus the eigenvalue spectrum
of the ghosts cancels that from $ \delta A_{0} $ and a subset of the
spectrum from $ M^{GF} $. We can therefore ignore
$\delta A_{0} $, and we must beware that not all the eigenvalues
of $ M^{GF} $ correspond to physical states.

In 3+1 dimensions, where the modes can have a $z$ dependence, the ghosts
cancel the contributions to the energy from a linear combinations of $
\delta A_{0}$ and $\delta A_{z} $, and the spectrum built on the
unphysical eigenmodes of $ M^{GF} $.

\subsection*{Numerical method}

We expand the scalar field perturbations in angular momentum states and
the gauge field in total angular momentum
states given by
\begin{displaymath}
A_{+} = \frac{e^{-i \theta } }{\sqrt{2} } (A_{r} - i \frac{A_{\theta }
}{r} ) \hspace{6mm} \mbox{ and } \hspace{6mm}
A_{-} = \frac{e^{i \theta } }{\sqrt{2} } (A_{r} + i \frac{A_{\theta }
}{r} )
\end{displaymath}
The background string solution is
\begin{displaymath}
\phi_{c} = f e^{iN \theta } ,\hspace{6mm}  A^{c}_{+} =
\frac{iNa}{\sqrt{2} r} e^{-i \theta },
\hspace{6mm} A^{c}_{-} = \frac{-iNa}{\sqrt{2} r} e^{i \theta },
\end{displaymath}
and the perturbations are
\begin{eqnarray}
\delta \phi  = \sum_{m} s_{m} e^{i(N+m) \theta }  &\qquad &
\delta \phi^{\ast }  =  \sum_{m} s_{-m}^{\ast } e^{-i(N-m) \theta }
\nonumber  \\
\delta A_{+}   =  \sum_{m} ia_{m} e^{i(m-1) \theta } &\qquad &
\delta A_{-}  =  -\sum_{m}ia_{-m}^{\ast } e^{i(m+1) \theta }.
\nonumber
\end{eqnarray}
Substituting in the above gives the eigenvalue equations
\begin{eqnarray}
\left( \begin{array}{cccc}
       D_{1} & V & A & B \\
       V & D_{2} & B & A \\
       A & B & D_{3} & 0 \\
       B & A & 0 & D_{4}
       \end{array}
       \right)
       \left( \begin{array}{c}
              s_{m} \\ s_{-m}^{\ast } \\ a_{m} \\ a_{-m}^{\ast }
              \end{array}
              \right)
              & = & \omega^{2} \left(
              \begin{array}{c}
              s_{m} \\ s_{-m}^{\ast } \\ a_{m} \\ a_{-m}^{\ast }
              \end{array}
              \right), \nonumber
\end{eqnarray}
where
\begin{eqnarray}
D_{1} & = & -\nabla^{2} + \frac{(N(1-a) + m)^{2} }{r^{2} } + \beta
(2f^{2} -1 ) + f^{2} \nonumber \\
D_{2} & = & -\nabla^{2} + \frac{(N(1-a) - m)^{2} }{r^{2} } + \beta
(2f^{2} -1 ) + f^{2} \nonumber \\
D_{3} & = & -\nabla^{2} + \frac{(m-1)^{2} }{r^{2} } + 2f^{2} \nonumber
\\
D_{4} & = & -\nabla^{2} + \frac{(m+1)^{2} }{r^{2} } + 2f^{2} \nonumber
\\
V     & = & (\beta -1 )f^{2} \nonumber \\
A     & = & \sqrt{2} \left( \frac{df}{dr} + \frac{Nf}{r} (a-1) \right)
\nonumber \\
B     & = & - \sqrt{2} \left( \frac{df}{dr} - \frac{Nf}{r} (a-1)
\right). \nonumber
\end{eqnarray}
As before, the eigenvalue equations are the same after complex
conjugation and changing $ m \rightarrow -m $
so we need consider only $ m \geq 0 $.

If we resolve the functions $ s_{m} $ , $ s_{-m}^{\ast } $ , $ a_{m} $ ,
$ a_{-m}^{\ast } $ into real and
imaginary parts then, as for the global string, the complex eigenproblem
separates into two eigenproblems
with explicitly real fuctions
\begin{eqnarray}
\left( \begin{array}{cccc}
       D_{1} & V & A & B \\
       V & D_{2} & B & A \\
       A & B & D_{3} & 0 \\
       B & A & 0 & D_{4}
       \end{array}
       \right)
       \left( \begin{array}{c}
              s_{m}^{r} \\ s_{-m}^{r} \\ a_{m}^{r} \\ a_{-m}^{r}
              \end{array}
              \right)
              & = & \omega^{2} \left(
              \begin{array}{c}
              s_{m}^{r} \\ s_{-m}^{r} \\ a_{m}^{r} \\ a_{-m}^{r}
              \end{array}
              \right) \\
\left( \begin{array}{cccc}
       D_{1} & V & A & B \\
       V & D_{2} & B & A \\
       A & B & D_{3} & 0 \\
       B & A & 0 & D_{4}
       \end{array}
       \right)
       \left( \begin{array}{c}
              s_{m}^{i} \\ -s_{-m}^{i} \\ a_{m}^{i} \\ -a_{-m}^{i}
              \end{array}
              \right)
              & = & \omega^{2} \left(
              \begin{array}{c}
              s_{m}^{i} \\ -s_{-m}^{i} \\ a_{m}^{i} \\ -a_{-m}^{i}
              \end{array}
              \right)
\end{eqnarray}
where $ D_{\alpha} $ , $ V $ , $ A $ and $ B $ are given above.

First, consider the vacuum, where $ f=1 , N=0 $ and $ A_{i} =0 $. Then $
A=B=0 , D_{1} =D_{2} $ in the above,
and the upper left block can be diagonalised with the eigenvectors
$ (1,1)^{T} s_{m} $ and $ (1,-1)^{T} s_{m} $ to give the four Bessel's
equations
\begin{eqnarray}
\left( -\nabla^{2} + \frac{m^{2} }{r^{2} } + 2\beta \right) s_{m} & = &
\omega^{2} s_{m} \\
\left( -\nabla^{2} + \frac{m^{2} }{r^{2} } + 2 \right) s_{m} & = &
\omega^{2} s_{m} \\
\left( -\nabla^{2} + \frac{(m-1)^{2} }{r^{2} } + 2 \right) a_{m} & = &
\omega^{2} a_{m} \\
\left( -\nabla^{2} + \frac{(m+1)^{2} }{r^{2} } + 2 \right) a_{-m} & = &
\omega^{2} a_{-m}.
\end{eqnarray}
The eigenvalue spectra of these are $ \omega^{2} = k^{2} + 2\beta $ for
(12) corresponding to that of a Higgs
particle, and $ \omega^{2} = k^{2} + 2 $ for the other three
corresponding respectively to the longitudinal and transverse states of
the massive gauge boson.

In the presence of the string background we still obtain the continua
given above, but as for the global string,
there are discrete eigenvalues correponding to particle states trapped
on the string.
The profiles $ f(r) $ and $ a(r) $ were solved for with a
relaxation method on the energy functional, and then substituted into
the
eigenvalue equations above. The eigenproblem was solved with the
boundary conditions $ s_{m} , s_{-m} , a_{m},
a_{-m} \rightarrow 0 $ as $ r \rightarrow \infty $ for various values of
$ m $, by the same method used for the global string. This was
performed for linearly descretized lattices with 50 and 100 points. The
results given in Table 3 are for 100
points and the translation modes  give an estimate of the accuracy. Note
that the error in the translation
modes is greater for $ \beta =4 $ than $ \beta =1 $ or 0.25. This is
because the
Higgs core is narrower with a much sharper transition from 0 to 1 than
in the other cases, and so there are fewer
points describing this region. One might think that a similar situation
would hold for the gauge field for
$ \beta =0.25 $ but its transition from 0 to 1 is more
gentle.

\subsection*{Zero modes}

After gauge fixing the only mode which can have zero eigenvalue is the
translation mode. The field configurations
corresponding to a gauge fixed infinitesimal translation are
\begin{eqnarray}
\phi (x) \rightarrow \phi (x + \delta x) & = & \phi (x) + (D_{j} \phi )
\delta x^{j} \nonumber \\
   & = & \phi + \epsilon \delta \phi \nonumber \\
A_{i} (x) \rightarrow A_{i} (x + \delta x) & = & A_{i} (x) + F_{ji}
\delta x^{j} \nonumber \\
   & = & A_{i} + \epsilon \delta A_{i}. \nonumber
\end{eqnarray}
These perturbations satisfy the gauge condition (9) by the equations of
motion. They also satisfy
\begin{displaymath}
M \delta \Phi  = 0.
\end{displaymath}
by the equations of motion. The numerical results for the translation
eigenvalues of $ M^{GF} $ are given in
Table 3.

\subsection*{Bound states}

Let us first consider $ m=0 $. In this case $ D_{1} =D_{2} $ and $ D_{3}
=D_{4} $, and $ M $ has eigenvectors of the form
$ (s_{0},s_{0},a_{0},a_{0})^{T} , (s_{0},-s_{0},a_{0},-a_{0})^{T} $.
Hence there are two sets of solutions for $ s_{0} $ and $ a_{0} $,
each  satisfying one of two equations:
\begin{eqnarray}
\left( \begin{array}{cc}
       D_{1} + V & A + B \\
       A + B & D_{3}
       \end{array}
       \right)
       \left( \begin{array}{c}
              s_{0} \\ a_{0}
              \end{array}
              \right)
              & = & \omega^{2} \left(
              \begin{array}{c}
              s_{0} \\ a_{0}
              \end{array}
              \right) \label{eBSHig}\\
\left( \begin{array}{cc}
       D_{1} - V & A - B \\
       A - B & D_{3}
       \end{array}
       \right)
       \left( \begin{array}{c}
              s_{0} \\ a_{0}
              \end{array}
              \right)
              & = & \omega^{2} \left(
              \begin{array}{c}
              s_{0} \\ a_{0}
              \end{array}
              \right).\label{eBSGau}
\end{eqnarray}
For each $ N $, we find that there exists a maximum value of $ \beta $
below which equation (\ref{eBSHig}) possesses one bound state solution,
corresponding to some sort of coupled Higgs-gauge boson trapped on the
string. For the $ N=1 $ string the bound state exists for
$ \beta <1.5 $. For $ \beta > 1.5 $ this bound mode becomes one of a
number of modes in the continuum, for which
the Higgs field is bound but the gauge field is not. These modes
correspond to some sort of resonant scattering
state of the gauge field. For $ \beta <1 $ there are similar
resonant scattering modes, but with the Higgs and gauge fields
interchanged.
For higher values of $ N $ there is still only one bound mode, but the
upper limit on $ \beta $ is higher. There
are also several resonant scattering states.

The solutions to equation (\ref{eBSGau}) also possess bound states for a
range of $ \beta $. For the range $ 1 < \beta < 8 $ there is
one such mode. For $ \beta <1 $, and for higher $ N $, where there are
a number of bound states. By examining the asymptotic form of
the perturbation operator, one can establish that the mass
squared is 2, and thus we are dealing with modes of the gauge field.
The gauge mode is trapped on the string because it has a lower mass
inside the Higgs core, so we would
expect the bound mode to have lower energy and for there to be more of
them when the Higgs core is wider than the
gauge field core. This is indeed the case.  However, as we warned
earlier,
gauge field modes are not necessarily physical.  In fact, we find that
these vector bound states have very similar eigenvalues to the
ghost modes, strongly indicating that they are just
a result of the residual gauge invariance in the background gauge.

Table 3 shows the eigenvalues of the lowest bound modes occuring for
equation (\ref{eBSHig}), equation (\ref{eBSGau}) and
for the ghosts. One can see that the ghost bound modes eigenvalues are
very close to those from  equation (\ref{eBSGau}).  The zero mode
eigenvalues give an indication of the error:  the difference between
the gauge field and the ghost eigenvalues is actually greater than
the departure of the numerical zero mode eigenvalue, but we
believe that our identification of them is nevertheless correct.

In 3+1 dimensions, there is also the $A_z$ field to consider.  This
has exactly the same perturbation equation as the ghosts and
$\delta A_0$. As there are in total 2 ghost degrees of freedom, they can
only cancel one linear combination of $\delta A_0$ and $\delta A_z$,
leaving the other for real physical propagating modes, at least one of
which is a bound state for $\beta < 8$.  Thus we have a massive gauge
boson
confined to the string.

\subsection*{Decay modes}

Abelian-Higgs strings with $ N >1 $ for $ \beta > 1 $ are unstable to
decay to strings of lower winding number.
The gauge field present in the local strings has the effect of removing
the energy's logarithmic dependence on
the radius, and so the allowed decays of local strings are $ N $ wound
string $ \rightarrow  M $ strings of winding
$ M_{i} $ with $ \sum_{i} M_{i} = N $ and $ M_{i} \geq 1 $. For $ \beta
< 1 $ the modes are the same but the
eigenvalues become positive. Consider the eigenvalues given in Table 3
for the $ m=2 $ mode for $ N=2 $.
When $ \beta =4 $ the string is unstable to splitting into two $ N=1 $
strings. As is well known $ \beta =1 $
strings do not interact and so $ \omega^{2} =0 $ for this case. Now when
$ \beta = 0.25 $ this splitting mode
has a small positive eigenvalue. This mode then corresponds to two $ N=1
$ strings oscillating about their
center of mass position. Again if we consider the $ z $ direction we
obtain travelling waves along the string.
Table 3 gives the eigenvalues of the splitting modes for various $
\beta $.

\section*{Discussion}

For global strings of unit winding number, we have found a bound mode
with zero angular momentum which, when quantised, corresponds to a bound
Higgs particle. The interaction terms in the
lagrangian, however, give a vertex for the Higgs boson to decay to two
Goldstone bosons, with a decay rate $ \Gamma \sim m_{H} $.
A Higgs boson occurs as a resonance in the scattering of two Goldstone
bosons at a centre of mass energy $ m_{H}^{2} $.
Similarly, the Higgs boson trapped on the string will be seen as a
resonance at a centre of mass energy of approximately $ 0.81 m_{H}^{2}
$.  Indeed, in a strongly coupled theory, the state could only be seen
as a very broad resonance, rather like the $\sigma$ in $\pi\pi$
scattering.

For local strings we have found a bound coupled Higgs-gauge boson for $
\beta <1.5 $ and a bound gauge boson in 3+1 dimensions for
$ 1< \beta <8 $. There are vertices for a Higgs boson to decay to gauge
bosons, but for the bound modes this cannot occur due
to energy conservation. So for $U(1)$ cosmic strings,  there are bound modes
for the above $ \beta $ ranges. Populations of these bound states could
have interesting effects: if they are truly stable, they could alter the
equation of state of the string, adding to the mass density and
subtracting from the tension.  However, in a realistic model it is
rather unlikely that the string fields will not be coupled to light
fields, which offer a decay route to the bound states.

Global strings possess a condensed matter analogue in the
Ginzburg-Pitaevski theory \cite{G-P,T-T} of superfluid Helium
II. In the G-P theory
the free energy of the condensate is given by
\begin{displaymath}
F= \int d^{3} x \left( \frac{\hbar^{2} }{2m_{4} } | \nabla \psi |^{2} -
\alpha | \psi |^{2} + \frac{\beta }{2} |\psi |^{4} \right),
\end{displaymath}
where $ \psi $ is the condensate wave function, which for stationary
superfluid has $ | \psi_0 |^{2} = \rho_{s} /m_{4} = \alpha /\beta $,
where $ \rho_{s} $ is the superfluid density and $ m_{4} $ is the mass
of a helium atom.
The equation of motion is first order in time, commensurate
with the non-relativistic nature of the system:
$$
i\hbar\frac{\partial \psi}{\partial t} =
\frac{\delta F}{\delta \psi^*},
$$
There exists an extremum of the free energy
of the form $ f(r) e^{iN\theta } $ where $ f(r) $ is the same as in
Section 1. This solution decribes a vortex in the superfluid in
which the superfluid flows at velocity $ v_s = \hbar N /m_4 r $ about a
line along the $ z $-axis, where the superfluid density is
zero. This is analogous to the global cosmic string. Unfortunately,
we cannot rely too heavily on this description of
vortices in Helium II, because the core size is of the same order
as the inter-atomic spacing -- the theory is strongly coupled.
It is nevertheless interesting to expand about the vortex
solution to find the spectrum of states in the vortex background.
To make contact with the dimensionless version of the perturbation
matrix $M$ of Section 2, it is convenient to scale all distances
in units of the correlation length $\xi = \surd(\hbar^2 /2m_4 \alpha)$,
and to use a dimensionless field $\chi = \psi \surd(\beta/\alpha)$.
Then for perturbations $\delta \chi$ we find
\begin{displaymath}
 i \hbar \frac{\partial}{\partial t}
\left( \begin{array}{c}
        \delta \chi \\ \delta\chi^*
       \end{array}
\right) =
\alpha (|\psi_0|^2 \xi^3) M
\left( \begin{array}{c}
        \delta \chi \\ \delta\chi^*
       \end{array}
\right).
\end{displaymath}
Thus the energy eigenvalues in the matrix background of states
with momentum $\hbar k_z/\xi $ along the string are
$$
\varepsilon = \alpha ( |\psi_0|^2 \xi^3) (\omega^2 + k_z^2).
$$
Of particular interest is the spectrum of bound states at
$\omega^2 \simeq 0.81$,
which we believe has not been convinvingly established
before in Helium II vortices.  These are not the ``bound''
states found by Pitaevski \cite{Pit61}, which are modes with $m=1$
associated with the translation of the vortex as a whole.
What we are discussing is a zero angular momentum
oscillation of the size of the vortex -- a ``sausage'' mode.
Bound states at angular momentum 0 and 2
have been reported by
Grant \cite{Gra71}, but the accuracy of the numerics is open to
question, for both states were found very close to threshold
($\omega^2 = 1$).  We find no evidence of bound states at
$m=2$, and at $m=0$ $\omega^2$ is approximately 20\% below
threshold.

These bound modes may be observable as a resonance in the scattering of
phonons or second sound off the core of the vortex, but in the
light of the comments above about the applicability of G-P theory to
such a vortex, this may not be observable.

The Abelian-Higgs model also has a condensed matter analogue in the
Ginzburg-Landau theory \cite{G-L}
of superconductivity. It has been shown that there is a range of
conditions for which G-L theory is applicable \cite{Gork}.

In G-L theory the free energy is given by
\begin{displaymath}
F=\int d^{3} x \left( \frac{\hbar^{2} }{2m} |(\nabla_a - \frac{2ie}{\hbar
} A_a )\psi |^{2} + \frac{B_a^{2} }{2\mu_{0} }
- \alpha |\psi |^{2} + \frac{\beta }{2} |\psi |^{4} \right),
\end{displaymath}
where $ \psi $ is condensate wave function and $ B $ is the magnetic
field. For the condensate in the absence of magnetic fields
$ |\psi_0 |^{2} = \alpha /\beta $.
There are two length scales to use in constructing a dimensionless
free energy functional, the correlation length $\xi$, and the
penetration depth $\Lambda = \surd(\beta m/4e^2\mu_0\alpha)$.
The
Ginzburg-Landau parameter is defined as the ratio $\kappa = \Lambda/
\xi$.  Defining a dimensionless gauge field $v_i = 2e\Lambda A_i/
\hbar$, we find
\begin{displaymath}
F= \frac{1}{2\kappa^2} \alpha (|\psi_0|^2 \Lambda^3)
\int d^{3} y \left( |(\nabla_a - iv_a
)\chi |^{2} + \frac{1}{2}(\nabla\times \mbox{\boldmath $v$})^2 -
2\kappa^{2} |\chi |^{2}
+ \kappa^{2} |\chi |^{4} \right).
\end{displaymath}
We see that the Ginzburg-Landau parameter is analogous to
$ \surd(\beta/2) $,
where $\beta$ is the ratio of scalar to
gauge couplings used earlier. This gives the same
perturbation matrix $ M $ (and $ M^{GF} $), up to an overall
scale factor, and so
the energy eigenvalues of the states in the superconductor vortex
background are
$$
\varepsilon = \frac{1}{2\kappa^2} \alpha(|\psi_0|^2 \Lambda^3)
(\omega^2+k_z^2).
$$
The bound modes on the
string exist for a range of $\kappa$. Converting the $ \beta $
limits into $ \kappa $ gives that for $ 1/\sqrt{2} < \kappa < \sqrt{3}
/2 $ there are bound modes of type (a) (coupled Higgs-vector
modes), while for
$ 1/\sqrt{2} < \kappa < 2 $ there are bound modes of type (b)
(longitudinal gauge field modes).

Again, the observability of such modes is an open question,
complicated by the fact that superconductor vortices are already
known to have quasi-particle bound states \cite{CarDeGMat63}.

\subsection*{Acknowledgements}
We are grateful to Chris Jones and David Waxman for helpful
conversations.  This work was supported by PPARC: MG by studentship
number 93300941 and MH by Advanced Fellowship number B/93/AF/1642.

\newpage

\section*{Table captions}

{\bf Table 1:} $ E_{Higgs} $ of global strings with winding number $ N $
in units $ \pi \eta^{2} $ and $m=0$ bound mode eigenvalues in units
of $ 2 \lambda \eta^{2}  $.
\newline
{\bf Table 2:} Eigenvalues of decay modes of global strings of winding
number $N$, in units of $2 \lambda \eta^{2}$. $m$ is the angular momentum
of the mode and the $M_i$ values are the winding numbers of the strings
formed by the decay.
\newline
{\bf Table 3:} Eigenvalues for local string modes in units of $ e^{2}
\eta^{2} /2 $. The modes shown are a)Higgs-gauge boson bound mode,
b)gauge boson bound mode, c)ghost bound mode t)translation mode,
and s)splitting modes.

\section*{Figure captions}

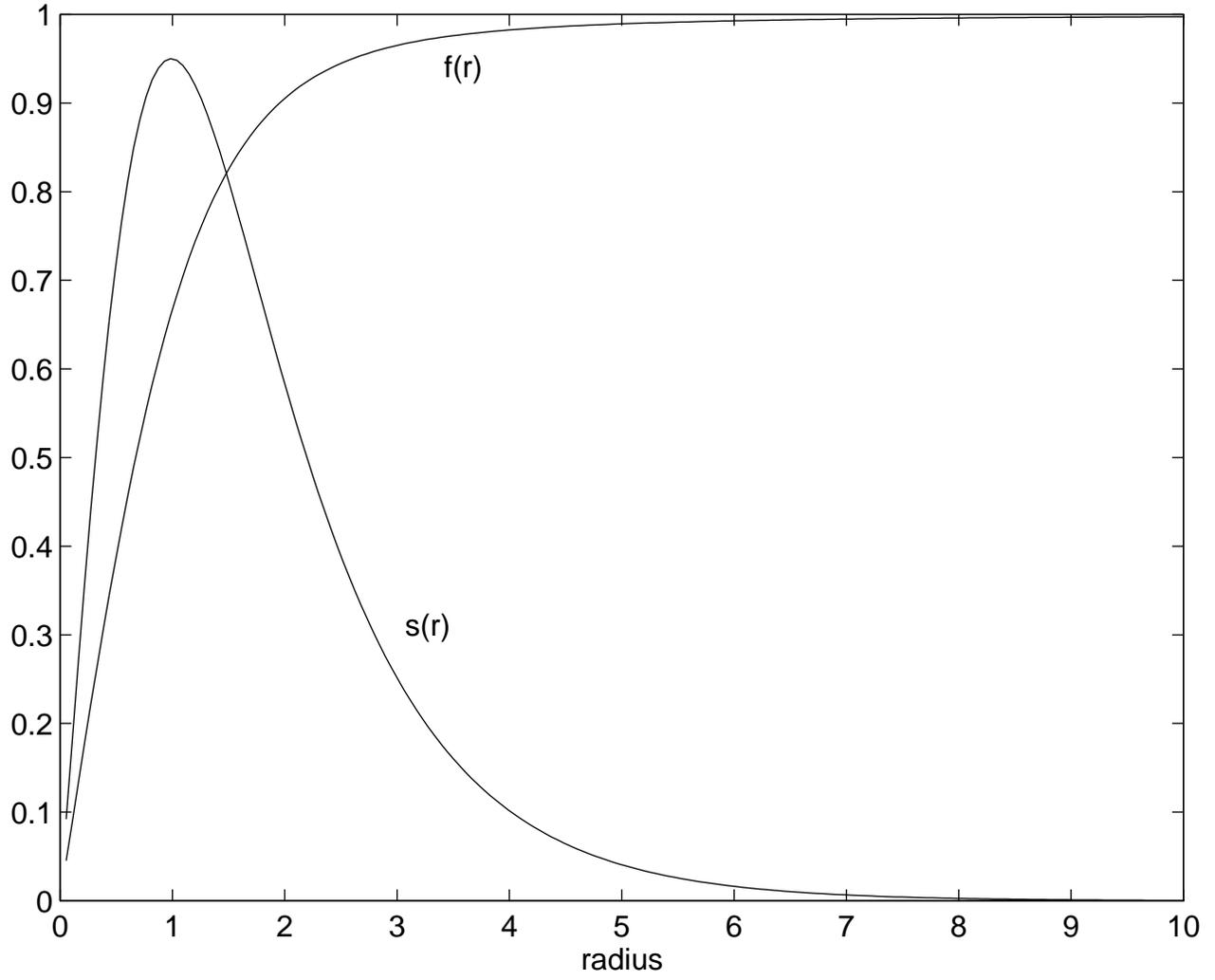
{\bf Figure 1:} Global string field profile $f(r)$ for $ N=1 $
and the $m=0$ bound mode field profile $s_0 (r)$.
\newline
{\bf Figure 2:} $ m=2 $ decay mode for a $ N=2 $ global string.
\newline
{\bf Figure 3:} $ m=2 $ decay mode for a $ N=3 $ global string.
\newline
{\bf Figure 4:} $ m=3 $ decay mode for a $ N=3 $ global string.
\newline
{\bf Figure 5:} $ m=4 $ decay mode for a $ N=3 $ global string.
\newline
{\bf Figure 6:} The energy per vortex for $ N=1, \ldots ,5 $ local
strings in units of $\pi \eta^2$.

\newpage

\begin{table}
\centering
\begin{tabular}{|l|l|l|lllll|}  \hline
N & $ E_{Higgs} $ & zero mode & \multicolumn{4}{l}{$ \omega^{2} $ of m=0
bound mode } & \\ \hline
1 & 0.7833 & 0.0012 & 0.8134 &       &       &       &       \\
2 & 2.4149 & 0.0019 & 0.6299 & 0.9554 &       &       &       \\
3 & 4.9835 & 0.0013 & 0.5180 & 0.8747 & 0.9727 &       &       \\
4 & 8.5712 & 0.0013 & 0.4437 & 0.7926 & 0.9314 & 0.9822 &       \\
5 & 13.0310 & 0.0014 & 0.3907 & 0.7206 & 0.8835 & 0.9527 & 0.9976 \\
\hline
\end{tabular}
\caption{}
\end{table}

\begin{table}
\centering
\begin{tabular}{|l|l|l|r|lr|}  \hline
N & m & $ M_{i} $ & $ \sum_{i} M_{i}^{2} $ & $ \omega^{2} $ &   \\
\hline
2 & 2 & 1,1              &  2 & -0.2093 &  \\
3 & 2 & 1,1,1            &  3 & -0.1283 &  \\
  & 3 & 1,1,1            &  3 & -0.3289 &  \\
  & 4 & 1,1,1,1,-1       &  5 & -0.1207 &  \\
4 & 2 & 1,2,1            &  6 & -0.0889 &  \\
  & 3 & 1,1,1,1          &  4 & -0.2429 &  \\
  & 4 & 1,1,1,1          &  4 & -0.3889 & -0.0617  \\
  & 5 & 1,1,1,1,1,-1     &  6 & -0.2410 &  \\
  & 6 & 1,1,1,1,1,1,-2   & 10 & -0.0749 &  \\
5 & 2 & 1,3,1            & 11 & -0.0662 &  \\
  & 3 & 1,1,1,2          &  7 & -0.1908 &  \\
  & 4 & 1,1,1,1,1        &  5 & -0.3151 & -0.0369  \\
  & 5 & 1,1,1,1,1        &  5 & -0.4224 & -0.1643  \\
  & 6 & 1,1,1,1,1,1,-1   &  7 & -0.3145 & -0.0272  \\
  & 7 & 1,1,1,1,1,1,1,-2 & 11 & -0.1868 &  \\
  & 8 & 1,1,1,1,1,1,1,1,-3 & 17 & -0.0475 &  \\ \hline
\end{tabular}
\caption{}
\end{table}

\begin{table}
\centering
\begin{tabular}{|l|l|l|l|l|l|} \hline
N & m & mode & \multicolumn{3}{l|}{ $ \omega^{2} $ } \\  \hline
  &   &      & $ \beta =4 $ & $ \beta =1 $ & $ \beta =0.25 $ \\ \hline
1 & 0 & a    & ---          & 1.5540       & 0.4538      \\
  &   & b    & 1.9097       & 1.5302       & 1.1073      \\
  &   & c    & 1.9304       & 1.5565       & 1.1298      \\
  & 1 & t    & -0.0459      & -0.0187      & -0.0096     \\
2 & 0 & a    & ---          & 1.0766       & 0.3604      \\
  &   & b    & 1.5110       & 1.0476       & 0.6775      \\
  &   & c    & 1.5713       & 1.0781       & 0.6963      \\
  & 1 & t    & -0.0267      & -0.0050      & -0.0001     \\
  & 2 & s    & -0.7654      & -0.0007      & 0.1121      \\
3 & 0 & a    & 1.7085       & 0.8043       & 0.2941      \\
  &   & b    & 1.2163       & 0.7856       & 0.4860      \\
  &   & c    & 1.2720       & 0.8057       & 0.4932      \\
  & 1 & t    & -0.0152      & -0.0013      & 0.0016      \\
  & 2 & s    & -0.4059      & -0.0009      & 0.0617      \\
  & 3 & s    & -1.3497      & 0.0023       & 0.1952      \\ \hline
\end{tabular}
\caption{}
\end{table}

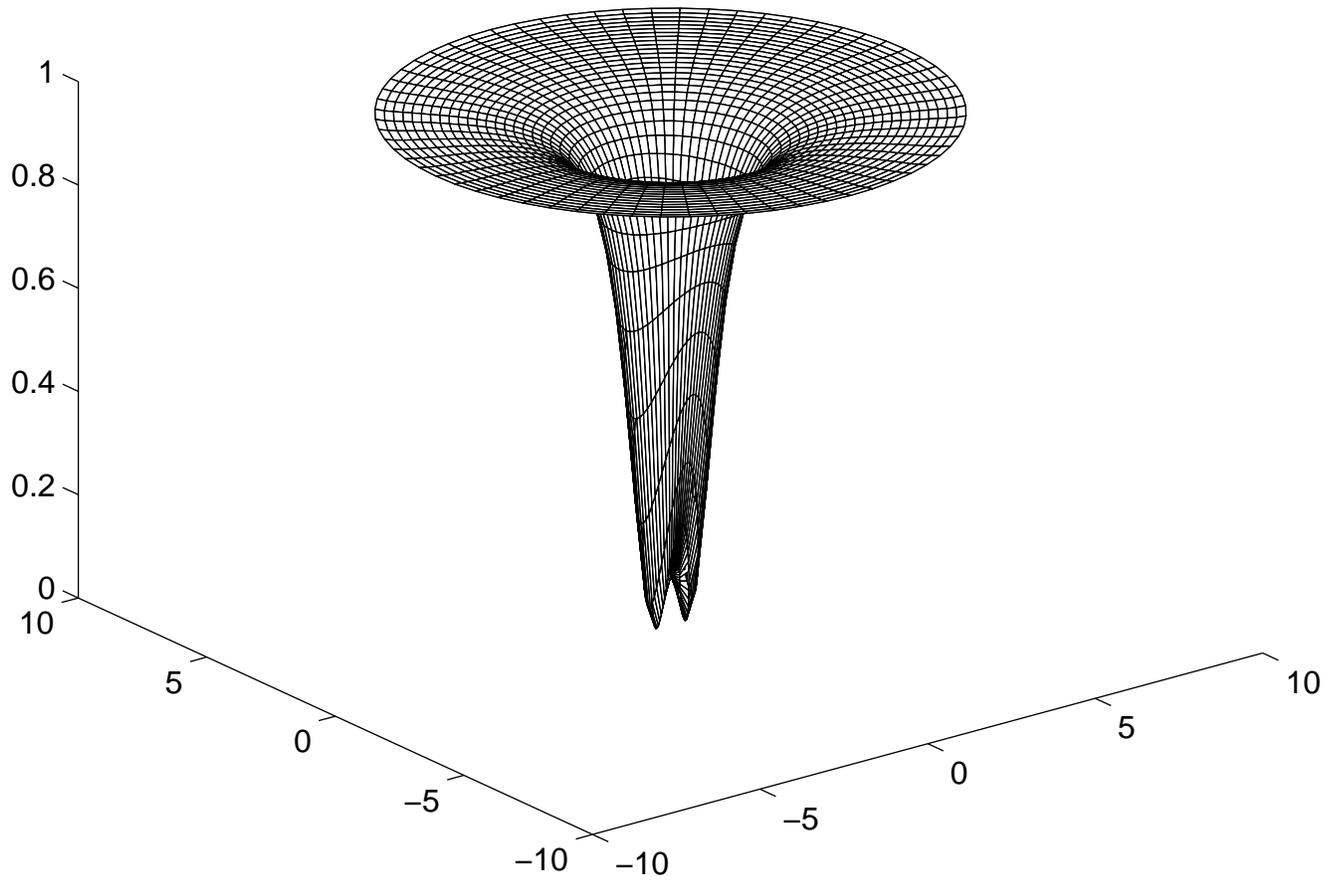
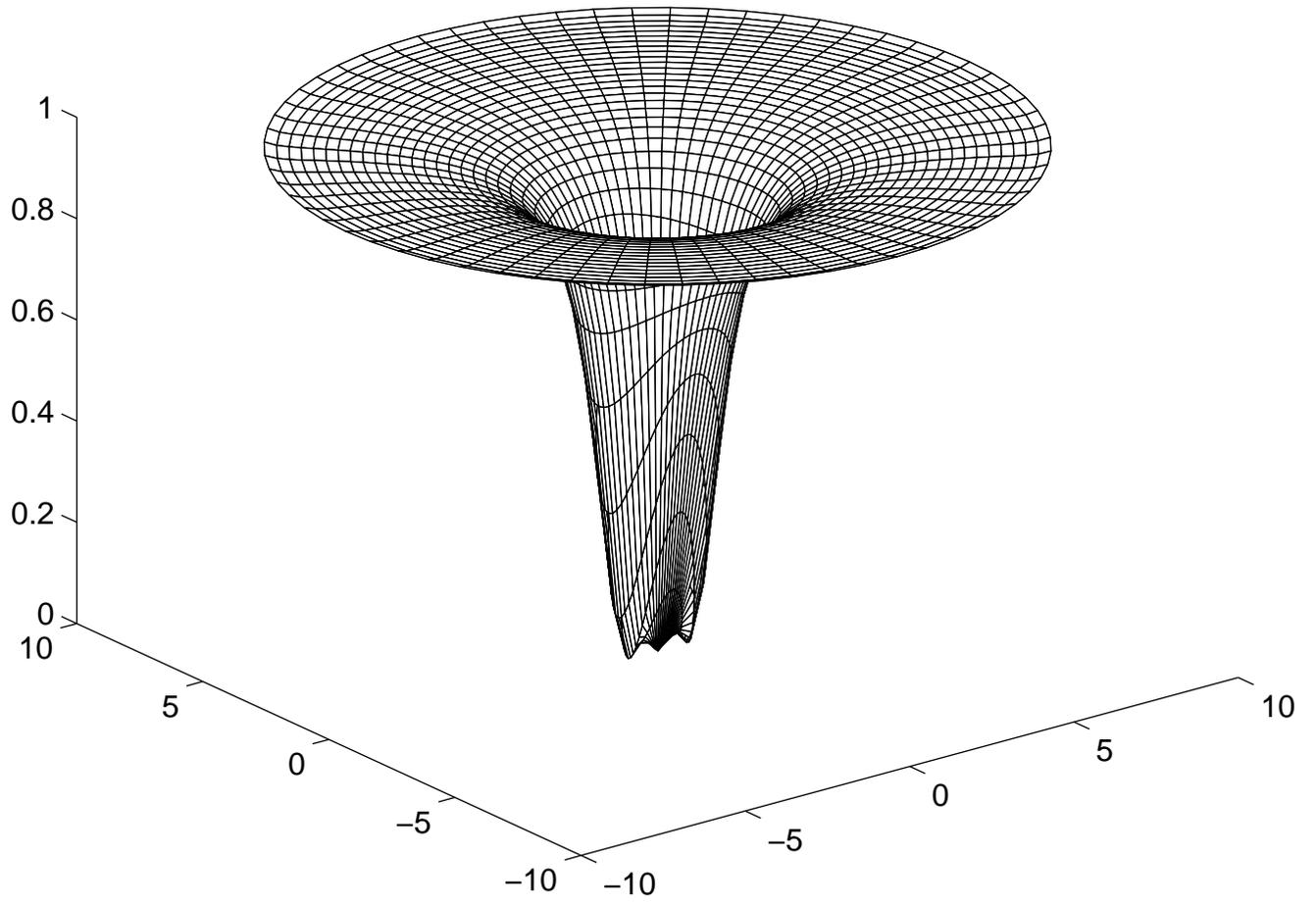
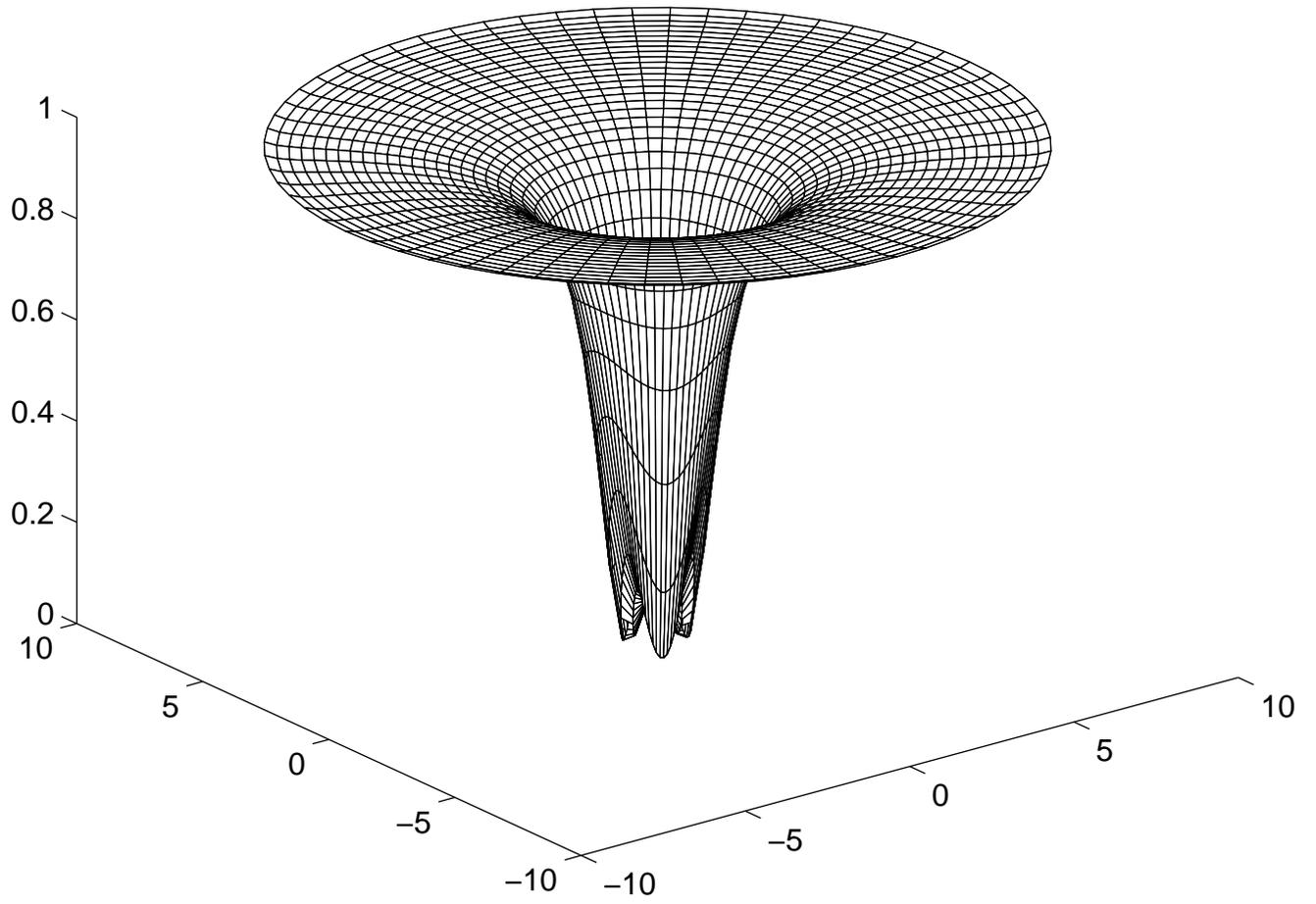
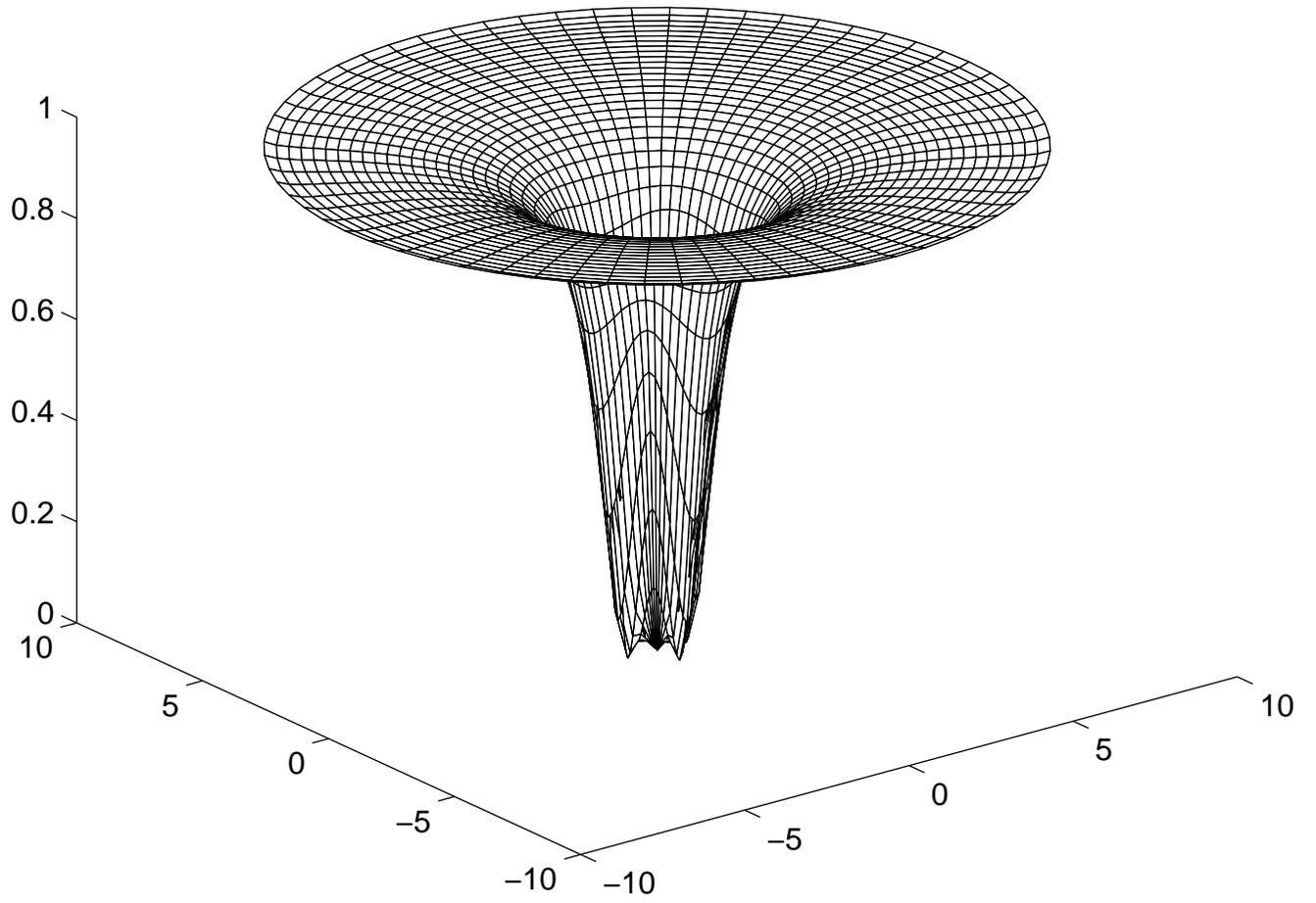
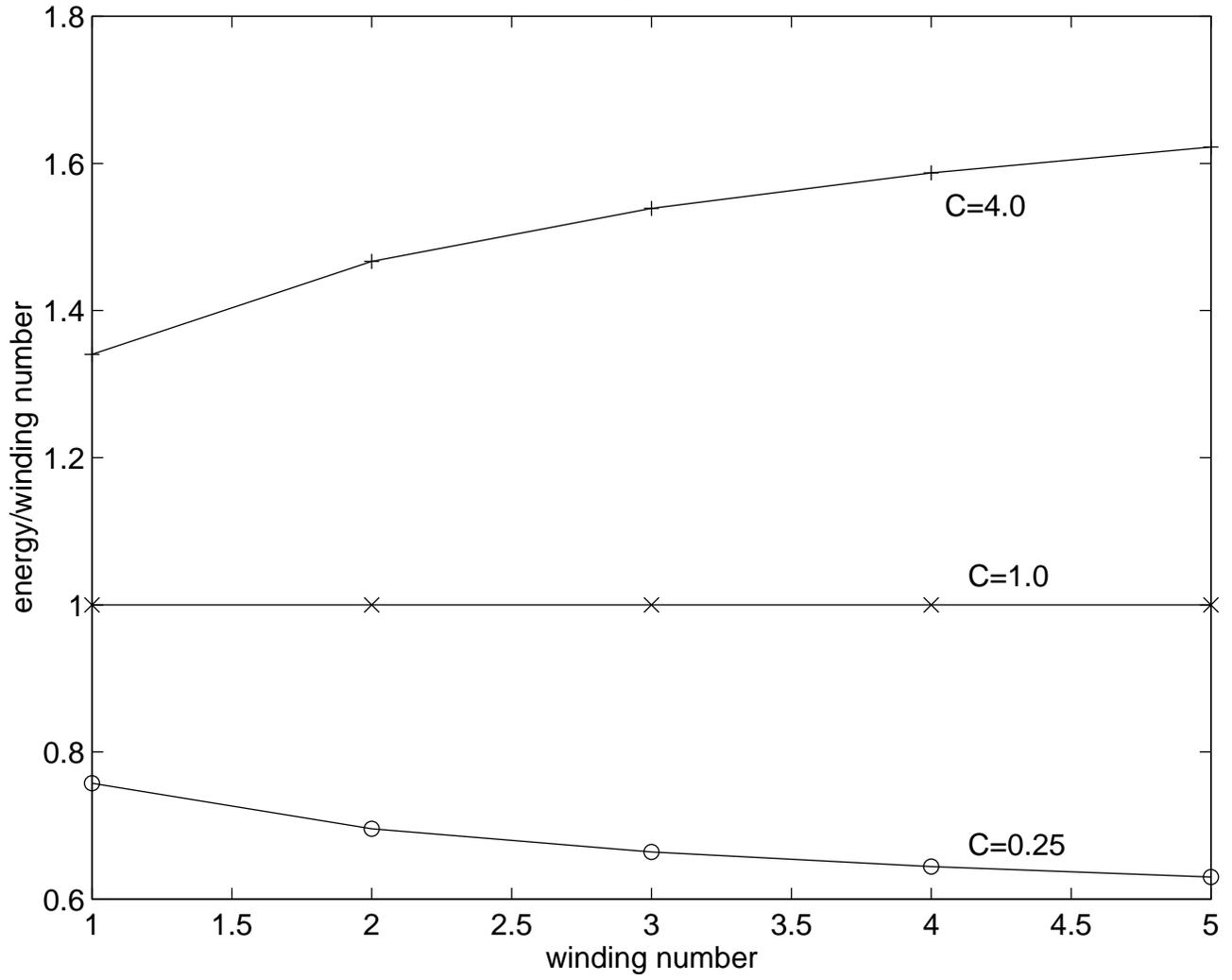
\end{document}